# LULC classification methodology based on simple Convolutional Neural Network to map complex urban forms at finer scale: Evidence from Mumbai

Deepank Verma[1]

Centre for Urban Science and Engineering

Indian Institute of Technology Bombay, Mumbai, India

Email: deepank@iitb.ac.in

Arnab Jana

Centre for Urban Science and Engineering

Indian Institute of Technology Bombay, Mumbai, India

Email: arnab.jana@iitb.ac.in

**Abstract.** The satellite imagery classification task is fundamental to spatial knowledge discovery. Several image classification methods are used to create standardized Land use and Land cover (LULC) maps, which facilitate research on spatial and ecological processes and human activities. Local Climate Zones (LCZ) classification maps are an example of standardized maps which have been widely used to demarcate the homogeneity in built and natural character in the cities. The LCZ classification scheme is primarily focused on urban climate-related research, in which 17 climate zones are mapped in a city area with the 100-150m spatial resolution. Each zone exhibits physical properties related to urban form and functions essential for thermal behavior studies. Extending this widely adopted approach to create LULC maps at finer resolution using the LCZ mapping scheme would benefit the allied domains of urban planning, transportation, and water resources management. This study proposes a novel solution to generate classification maps with a 10-band Sentinel-2B dataset and Convolutional Neural Networks (CNN) at the 10m spatial resolution. The classification benefits from CNN's property to preserve local structures in the image datasets. The proposed CNN model outperforms traditional machine learning models such as Artificial Neural Network, Random Forests, and Support Vector Machines. The overall accuracy and kappa statistic of the CNN model trained on 14 urban and natural classes are 82 percent and 0.81, respectively. The study also discusses the utility of the model for specialized remote sensing tasks such as change detection, identification of slum settlements, and mapping pervious/ impervious layers in urban settlements with higher accuracy.

## 1. Introduction

The LULC classification aims to achieve uniform categorization of landforms at various scales [1]. Such classification is an essential component in the creation of standardized maps which help in decision- and plan-making processes. LULC maps have been used to estimate agricultural production [2]–[4] , study urban change detection [5]–[7], climate [8], [9], biodiversity [10], [11] and to map natural hazards [12], [13]. Earth observation (EO) datasets, which have been made available by different space agencies, have fostered advancement in image classification methods. Parametric techniques such as Maximum Likelihood (ML), and non-parametric techniques such as Support Vector Machines (SVM), Random Forests (RF), and Artificial Neural Networks (ANN) have been extensively experimented with per-pixel image classification tasks [14]–[16]. Such widely adopted classification techniques have been challenged by spatial-spectral classification methods such as Object-Based image analysis (OBIA) which have shown significant improvement in classification accuracy [17]. However, the OBIA-based research has been restricted to the Very High/High resolution (VHR/HR) image datasets, which have limited availability to researchers. Among various EO datasets, the Landsat dataset series (Landsat 1-8) have been widely used in classification tasks. Due to the coarser spatial resolution of imagery products (Landsat 8: 15m (Pan) - 30m – 100m (Thermal bands)), these datasets have been widely utilized to study mainly regional characteristics with broad land cover classes such as built, croplands, water bodies, forests, and fallow lands [16]. In recent years, Landsat along with Sentinel products have been utilized to study intra-urban characteristics with improved classification techniques [18]–[20].

---

[1] Corresponding Author

One of the prerequisite components in any LULC classification study is the selection of a classification system. The classification system is usually designed to cover the user's requirement, availability of reference samples and classification algorithms, and reproducibility at various scales [21]. Some of the common land cover classification system used by organizations such as United Nations Educational Scientific and Cultural Organization (UNESCO) [22], Food and Agriculture Organization (FAO) [23], Federal Geographic Data Committee (FGDC) [24], Commission of the European Communities (CEC) [25] and other LULC classification systems such as Urban Atlas [26], National Remote Sensing Agency (NRSA) [27] are utilized for large scale image classification for land monitoring and decision making. Such schemes utilize different EO datasets for assessment of mainly vegetation cover and broad land cover classifications. As most of the land cover schemes are applicable at lower resolution satellite imagery, hence the urban features are represented as aggregated classes such as a "built cover" or "continuous/discontinuous urban fabric". Anderson [1] presented a hierarchical LULC classification system at various levels of assessments which can be prepared with the help of different EO datasets along with the Land use maps prepared through ground surveys. Similar classification systems [28] have been in practice for urban development and management. Some of the major classes used in such classifications are Residential, Commercial, Recreation, Transportation, and Industrial which are further subdivided into detailed sub-classes based on the arrangement of built structures, density, and functions. However, the associated definitions and rulesets to define such classes lies with the governing authority of the area in question, hence generalization of such classification schemes has not been possible. Due to the lack of strict class boundaries and definitions with majority of classification systems and the unavailability of a well-documented approach to generate LULC maps, researchers have considered individual-requirement based classes to identify intra-urban features. Studies included methods such as pattern recognition to identify spatial patterns [29], delineation of impervious surfaces built types using morphological features [18], [30], [31], and creation of land use maps using various EO datasets [19], [32]. The study of underlying spatial form and function is an important area of research in urban planning, design, and development. However, traditional approaches to prepare classification maps have been based on inconsistent and limited observations due to a lack of a standardized approach and classification system. This study utilizes the Local Climate Zone (LCZ) classification system which consists of well-delineated intra-urban classes and large documentation explaining the classification process for reproducibility and universal applicability.

The LCZ classification system has been widely used in urban climate studies to classify intra-urban built and natural features and to facilitate recognition of urban morphology and built arrangement. In short, LCZ is LULC for urban climate studies [33]. The LCZ classification scheme consists of 17 classes, of which 10 classes define urban character and seven natural character. LCZ classes categorize the packing of roughness features of built structures, openness and vegetation to study permeability and urban geometry. The LCZ classification is a method to create a global database of urban form and function in different cities for urban climate studies [9]. The maps are utilized by scientists studying heat island effects in various urban neighborhoods [34], [35]. The existing approaches to LCZ classification include (a) imagery-based, in which the EO datasets along with the training samples created by local experts are used to prepare maps with the help of various classification algorithms [36], [37], and (b) GIS-based approaches, which include decision and rulemaking through GIS maps comprising of geographical objects such as building footprints, heights, and roads to create LCZ maps [38], [39].

The preparation of imagery-based classification maps is done through World Urban Database and Access Portal Tools (WUDAPT) [40]. The method utilizes open source GIS plugins to execute classification algorithms. It requires help from local experts to create training samples of the selected city. Landsat data with RF classifier is used for image classification in open source SAGA GIS software. In addition to the detailed step-by-step methodology [40] for the creation of such maps, various studies have experimented with additional datasets and methods which include transferability of training samples to other cities [41], the usage of ASTER data along with Landsat data [42], and Sentinel-1 SAR data with multispectral Landsat data [43], [44].

Traditional image classification tasks utilize the pixel-based classification methods, which do not take into account the spatial information of neighboring pixels. LCZ classes comprise of built and natural roughness features which, with context-aware pixel-based classification techniques, may produce an improved representation of the physical characteristics of an area [45]. Contextual classification approaches utilize neighboring pixels to solve the problem of intraclass spectral variations [21]. However, the majority of commonly used classification techniques, including LCZ, still utilize only spectral variables for image classification tasks [46]. While contextual classification methods [47]–[49] have shown significant improvement in accuracy in hyperspectral datasets, the studies showing replication of the methods in EO datasets such as Landsat and Sentinel have been scarce. In recent years, Convolutional Neural Networks (CNN) have shown significant advancements in feature detection and image classification tasks. CNN models are useful in learning the representation of spatial and spectral variations in the satellite dataset. While the majority of the practical use cases of CNN based models are based on image and video analysis [50], some studies have modified and utilized this concept in remote sensing domain [51]–[53]. Similar to OBIA, the studies have been mostly focused on VHR/MR imagery types. LULC classification with the help of CNN has not been tested with Medium Resolution satellite imagery. The only exception is hyperspectral imagery where CNN models have outperformed traditional techniques in image classification [54]–[56].

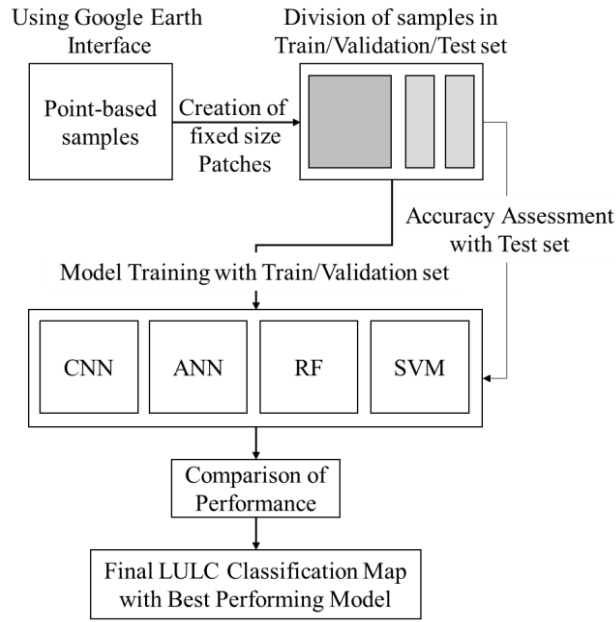

**Figure 1**: Flowchart showing methodology used in the study.

This study proposes a CNN based classifier for intra-urban built and natural areas classification at 10m spatial resolution. It utilizes 10-band Sentinel 2B satellite imagery to create training and testing datasets. This study further compares the classification results with other machine learning algorithms such as Artificial Neural Network (ANN), Random Forests (RF) and Support Vector Machines (SVM). These machine learning algorithms are modified to integrate spatial information post-classification for suitable comparison. Altogether, this study proposes a methodology to prepare the fine resolution classification map based on the LCZ classification scheme (Fig. 1). The created approach takes into account the spatial-spectral information to map complex urban features at higher details, which otherwise are difficult to classify with traditional classification techniques.

## 2. Selection of Area and dataset

The city of Mumbai is chosen to demonstrate the results of the created methodology. We collected the sentinel 2B dataset dated 15$^{th}$ March 2018 for this study. The Sentinel dataset comprises 13 spectral bands with varying spatial resolution (10, 20 and 60m). The dataset does not provide panchromatic band, which is commonly used to enhance the spatial resolution of bands with coarser resolution. With the help of 4 fine (10m) resolution bands (2, 3, 4 and 8), we pansharpened the remaining 20m resolution bands (5, 6, 7, 8A, 11 and 12). The process of pansharpening was followed from various studies [57], [58]. The resulting sentinel imagery consists of 10-bands each of 10m spatial resolution. Sentinel image dataset is represented in 16-bit data format, for which the normalization of data is done by dividing each Digital Number (DN) of the pixel with the maximum DN value of the dataset.

Table 1: Built and land cover types considered in the study. Adapted from [33].

| | Built and Land Cover types | Definitions | | Built and Land Cover types | Definitions |
|---|---|---|---|---|---|
| LCZ 2 | Compact mid-rise | Dense mix of buildings with 3-9 stories. Land cover is mostly impervious. | LCZ A | Dense Trees | Heavily wooded landscape of deciduous and/or evergreen trees. |
| LCZ 3 | Compact low-rise | Dense mix of buildings with 1-3 stories. Land cover is hard-packed and impervious. | LCZ B | Scattered Trees | Lightly wooded landscape of deciduous and/or evergreen trees. |
| LCZ 4 | Open High Rise | Open arrangement of tall buildings. Combination of arrangement with impervious and pervious land cover. | LCZ C | Bush, scrub | Open arrangement of bushes, shrubs and short woody trees. |
| LCZ 5 | Open mid-rise | Open arrangement of buildings with 3-9 stories. Abundance of pervious land cover. | LCZ D | Low Plants | Featureless landscape of grass and herbaceous plants/crops. |
| LCZ 8 | Large low-rise | Open arrangement of buildings with 1-3 stories. Impervious land cover. | LCZ E | Bare rock or paved | Featureless landscape of rock or paved cover. |
| LCZ 9 | Sparsely built | Sparse arrangement of small or medium sized buildings in a natural setting. Pervious land cover. | LCZ F | Bare soil or sand | Featureless landscape of soil or sand cover. |
| LCZ 10 | Heavy Industry | Low-rise and mid-rise industrial structures (tower, | LCZ G | Water | Large open water bodies such as seas and lakes, or small |

| | | |
|---|---|---|
| tanks, stacks). Impervious land cover. | | bodies such as rivers, reservoirs, and lagoons. |

By visual inspection of the city of Mumbai through high-resolution Google satellite imagery, many observations can be made. The city predominantly comprises of compact low rise (LCZ 3), and open mid-rise (LCZ 5) built arrangements in the form of squatter settlements and housing colonies respectively. High-rise buildings (LCZ 4), on the other hand, have scattered presence throughout the city. Most of the high-rise buildings are part of small townships surrounded by other natural and built classes making the distinction between such built structures difficult. Compact mid-rise (LCZ 2) settlements are dominated by Victorian-era built structures which are apparent in the southern part of the city. Mumbai city also consists of a considerable percentage of land cover under sparsely built (LCZ 9) dominated by academic and administrative institutions. LCZ 3 and Lightweight Low Rise (discussed in [33] as LCZ 7) classes differ in construction materials, the proper delineation between the two cannot be made through visual observation. Hence, these classes are collectively studied as LCZ 3. The presence of the rest of the classes (discussed in [33] as LCZ 1 and LCZ 6) is scarce and hence not considered in the study.

The natural classes in the city can be distinctly identified from the satellite map. The mangroves in the coastal areas of the city represent dense trees class (LCZ A), whereas the city forests and the national park in the city's northwestern part are categorized as scattered trees (LCZ B). Land cover comprising small shrubs and trees or plants between the built form and the parks or gardens can be classified as bush and scrub (LCZ C). Impervious surfaces such as roads, runways, and dockyards are identified as a paved class (LCZ E). The low plants (LCZ D) category includes parks, stadiums, playgrounds which have green vegetation cover. Bare soil or sand (LCZ F) class includes the sand in the beaches along the coastline, cleared forests or vegetation for built purposes, and salt pans. A total of 14 prominent classes in the city of Mumbai (Table. 1) are considered and samples are created with the help of the Google Earth platform.

The proposed methodology utilizes a 10-band Sentinel dataset. Similar to the image-based WUDAPT approach, training samples are collected with the help of Google Earth Pro software. However, instead of the creation of polygons, point-based samples are created [59]. For better generalization, all the areas of the city are considered for creating samples for each of the classes (Fig. 2). In this study, approximately 3500 points belonging to 14 classes are created which are further randomly split into the *train, validation,* and *test* sets in the proportion of 5:2:3. Patches of size 11x11x10 are extracted from point-based samples by creating a buffer of 5 pixels around each sample. Each patch is labeled using the central pixel of the patch.

**Table 2**: Point-based samples created for each class. Further divided into Train, Val and Test set in the ratio of 5:2:3.

| Classes | Total | Train | Val | Test |
|---|---|---|---|---|
| LCZ 2  | 253  | 116  | 65  | 72  |
| LCZ 3  | 498  | 242  | 118 | 138 |
| LCZ 4  | 321  | 148  | 77  | 96  |
| LCZ 5  | 237  | 115  | 41  | 81  |
| LCZ 8  | 180  | 80   | 38  | 62  |
| LCZ 9  | 159  | 92   | 21  | 46  |
| LCZ 10 | 207  | 102  | 46  | 59  |
| LCZ A  | 212  | 100  | 41  | 71  |
| LCZ B  | 258  | 124  | 56  | 78  |
| LCZ C  | 198  | 92   | 35  | 71  |
| LCZ D  | 253  | 116  | 72  | 65  |
| LCZ E  | 207  | 107  | 36  | 64  |
| LCZ F  | 416  | 232  | 75  | 109 |
| LCZ G  | 178  | 86   | 30  | 62  |
| | | | | |
| Total  | 3577 | 1752 | 751 | 1074 |

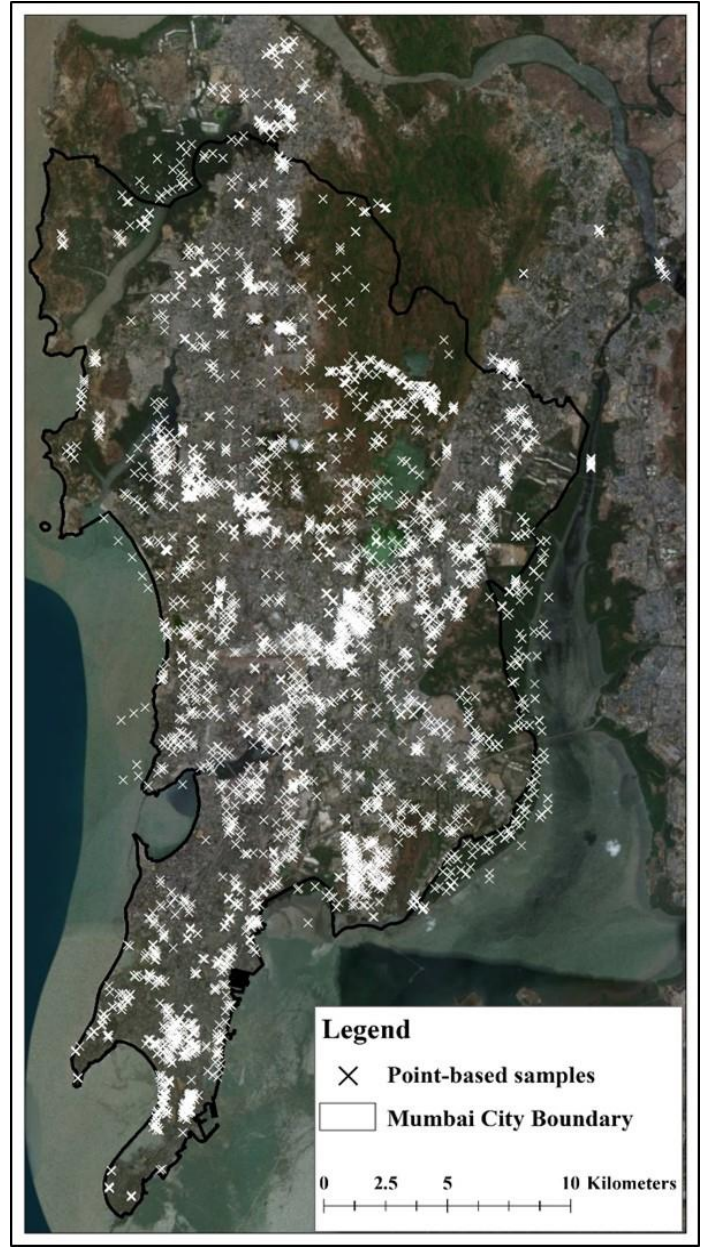

**Figure 2**: Map showing locations of selected point-samples.

Point-based training samples creation is more effective than the creation of polygons for two reasons. Firstly, in cities with spatial heterogeneity such as Mumbai, the majority of the considered LCZ classes coexist or overlap in a random fashion. Delineating such areas by polygon based training samples is difficult. For example, the scattered distribution of high rise clusters (LCZ 4) is predominant in the city; if drawn as polygons, this would result in smaller and unusable training samples. With the discussed approach, the buffer of fixed size is automatically generated over the samples which encompass the spatial neighborhood of the fixed size, thereby reducing the need for manually drawn polygons. Moreover, for further evaluation, the size of the neighborhood can be modified to select the good fit between the performance of the classifier and the neighborhood size. Secondly, CNN models expect a fixed input size, which is not practical with hand-drawn polygons. The equal size buffer created from the selected point samples ensure the consistency in the CNN model input.

# 3. Convolutional Neural Networks

Neural networks learn to detect patterns and representations from the datasets which are utilized in various classification tasks [60]. Neural networks consist of three main layers which comprise neurons (or nodes). The nodes in the input layer hold the data points. Each node is connected to the hidden layer which is further connected to the output layer. The activation functions are applied to hidden and output layers. Activation function introduces the non-linearity in the model by firing selected nodes in the layer. Every connection between the nodes has a weight. With every iteration, the loss is calculated with a loss function, and weights of each connection are modified with the backpropagation algorithm. The model is run until the loss is stabilized. The trained model is further used to predict the outcomes of the newer dataset. CNNs are an extension to regular neural networks. The input to the CNN models are the neurons arranged in the form of arrays of dimension N x M x R, where N, M, and R are the length, breadth and the depth. Typically, RGB images, which have a depth of 3, are used as inputs to the CNN model.

### 3.1. *CNN structure design: Network Structure*

The structure of CNN includes input layer, convolution layer, pooling layer, and fully connected layers. The designed CNN architecture (Fig. 3) consists of an input layer with dimensions of 11x11x10, which holds the individual image patches. A filter of size 3x3x10 is employed in the first convolution layer, which slides overall spatial locations and calculates the dot product of the filter and the small chunk of the input data (size similar to the filter), producing a 2-D activation map. 32 such activation maps are produced which are bundled to create a volume of 9x9x32. Activation functions are used at each convolutional layer to introduce the non-linearity in the model. RELU (Rectilinear Linear Unit) is commonly used in neural networks due to its better performance on loss convergence. RELU performs elementwise activation to the created volume of convolution layer. The filter of size 3x3x32 is applied to the second convolutional layer. 64 activation maps are created which produced the output volume of 7x7x64. Pooling is a downsampling operation, which helps in reducing the number of parameters in the network. Pooling operation is independently implemented for each activation layer. Max pooling with 3x3 filters and stride of 2 is applied to the output volume of the second convolution layer. The resulting volume after pooling becomes 3x3x64. The volume is flattened to create 576 values, which are fully connected to the array of 128-D vector. The vector is further connected to the 14-D vector which produces the predictions. The softmax function is applied to the final layer which provides the probabilities of the existence of each class. The CNN model is trained for 300 epochs.

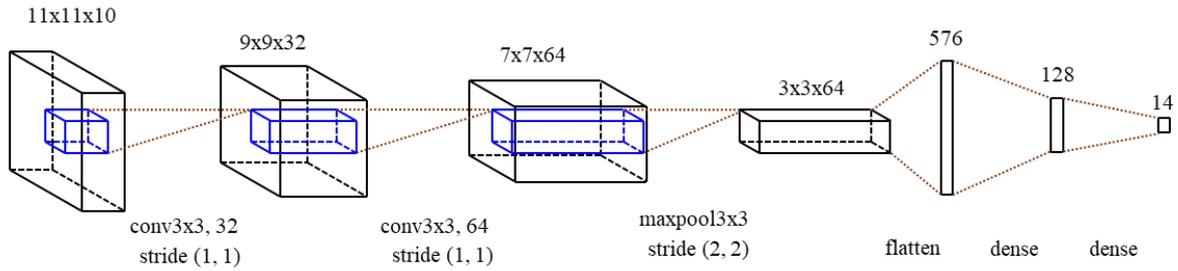

**Figure 3**: CNN Network Structure Diagram.

*3.2. Kernels*

The Kernel size is an important hyper-parameter which is responsible for creating a group of local regions in the provided input. The combinations of pixels in local regions provide a hierarchical understanding of the object present in the input. The CNN learns to classify shapes, colors and the distributions in an image. We compared the accuracy of the CNN model with different sizes of kernels (1x1, 3x3, and 5x5). The model showed the best performance when the size of the kernel is kept at 3x3.

*3.3. Data augmentation*

CNN, in general, require large datasets to provide satisfactory results. Data augmentation is a technique to manufacture the new data from the existing dataset without sufficient loss of representation of details. The random zoom, rotate, mirror, flip, contrast stretch, etc. are often applied to the images to generate more dataset. However, in the case of satellite image patches, most of the listed techniques would result in a loss of details. Therefore, we introduce the rotation factor to the *Train* and *Val* set, where each patch was rotated 90 degrees three times and saved. It led to the increment in the dataset utilized by the CNN classifier by the factor of 4 (2503 x 4 = 10012 samples).

*3.4. Dropout rate*

Dropout is a technique to reduce overfitting and improve generalization of the model [61]. The dropout rate is introduced in hidden layers as a probability factor at which the connections between the neurons are randomly omitted. This prevents too much reliance on the particular neurons. For example, for the dropout rate of 0.5, the model randomly drops the connections with 50% probability. The model is forced to learn the features of the input without relying on specific neurons, hence, it becomes better generalized to the dataset. We experimented with different values of dropout rates and at different layers. In this model, dropout rates of 0.5 and 0.25 are implemented at max pooling layer and the dense layer respectively.

4. **Other Machine Learning Classifiers**

Widely used machine learning algorithms such as RF, SVM and ANN in LCZ classification are further utilized to compare the per class accuracy of the CNN model. Random Forests (RF) is an ensemble of random decision trees, in which class assignments are merged to increase overall accuracy and stability in model predictions. Each decision tree will formulate a set of rules, which are used by the model in performing predictions. Random Forests model randomly subsets the input features and grows the decision tree from the calculated node. After the creation of n trees, the class labels are determined by considering the most voted class. RF classifiers are robust to noise, can handle high data dimensionality and insensitive to overfitting [15], [62]. We experimented with the RF classifier by varying the number of trees, before finally selecting 32 trees.

Given a labeled training dataset, Support Vector Machines (SVM) finds the best defining boundaries (an optimal hyperplane) which increase class separability of the n-Dimensional input features according to the class labels [14]. In a simpler two-dimensional dataset, the hyperplane is the line dividing the two dataset clusters. SVMs are capable of better generalization from relatively smaller datasets by performing complex feature transformations using the set of mathematical operations (known as Kernels) [16]. The classifiers are trained with the samples in patch based dataset (11x11x10 values), which are further subdivided into 121 independent features each with the size 1x10. The label of the particular patch is shared among the independent features.

An ANN with 2 hidden layers, each containing 20 nodes, is created in which the input layer holds the 10 feature values from each training sample. The output layer provides the predicted output with the help of the softmax classifier. The dropout rate of 0.5 is implemented in each hidden layer to prevent overfitting. The ANN model is trained for 250 epochs until the loss is diminished. Keras [63] python library is used to design CNN and ANN architectures, while Scikit-learn [64] is used to implement RF and SVM. The training, testing, and evaluation is done on a system with a quad-core Xeon processor with NVIDIA K2000 GPU card. The training time for each of the classifiers took less than 30 minutes.

*4.1. Integrating spatio-contextual information with spectral-only classifiers*

CNN, by design, incorporates spatial and spectral information for image classification. To compare the spectral-only classifiers with CNN based classification and to improve the performance of spectral-only classifiers, we incorporated spatial information in the image classification process at post-classification [65]. Commonly used techniques to include spatial context at pre-classification, during classification, and post-classification include utilization of image textures [66], Mathematical Morphology [67], Object-Based Image classification [68], Contextual support vector machines [49], Markov Random Fields [69], and Majority Voting [48]. In this study, Majority Voting (MV) method is utilized, which is considered accurate, simpler and faster than other spatial-spectral classification methods [48].

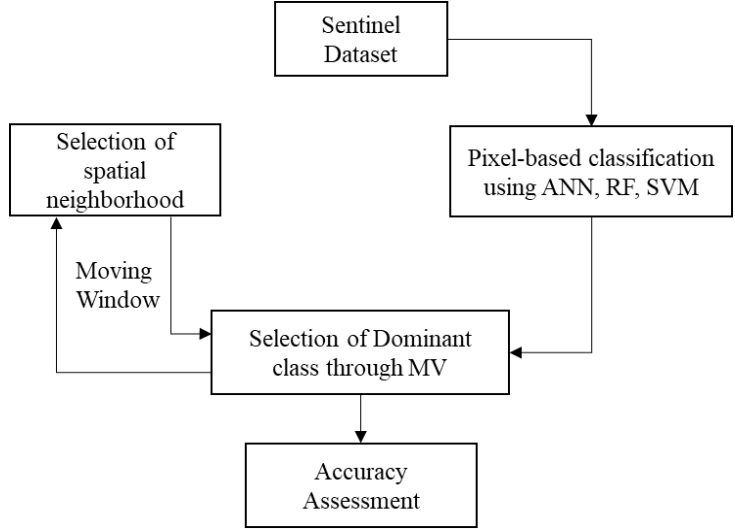

**Figure 4**: Integration of spatial information to pixel-based classification.

Majority Voting (MV) utilizes a spatial neighborhood prepared with the help of various segmentation algorithms such as Watershed [70], HSeg [71] and classification map as prepared by the spectral classifier [47]. For each segmented neighborhood, all the pixels are assigned to the most frequent class. This study performs a modified MV technique in which the fixed-size sliding spatial neighborhood (kernel) of size 11x11 pixels is utilized [45], which assigns the most frequent class to the central pixel (Fig. 4). For each pixel in the pixel-based classification map, the final output gets refined with MV-based spatial regulation. The accuracy statistics of all the three classifiers on the *Test* dataset is calculated with the help of final classification maps.

## 5. Results

We present the experimental results of the classifiers' performance on the *Evaluation* set. The *Train* dataset is used to train the model while *Val* dataset is used to provide an unbiased evaluation of the trained model while tuning hyperparameters to get the best performance of the developed Neural Network model. The *Test* set is used to provide an unbiased evaluation of the final prepared model. The *Test* set is not seen by the model during its training. It is used only after the model is fully trained after hyper-parameters tuning. We made use of *Train*, *Eval,* and *Test* set in case of CNN and ANN classifiers, in which hyperparameters such as kernels, dropout rate, hidden layers, depth of layers, and activation functions are used. However, for RF and SVM classifiers, we merged the *Train* and *Eval* set to create a larger training dataset and assessed the performance of the classifier on the *Test* dataset (Table. 2).

We utilized Overall Accuracy and Kappa metrics to compare the per class performance. We also calculated the F1-score metrics to compare the accuracy between classes. F1-score is a harmonic mean of Precision and Recall. Precision is the proportion of positive detections of the classifier which were actually correct, whereas Recall refers to the proportion of actual positives which were detected correctly. Precision is also defined as TP/TP+FP (TP: True Positives; FP: False Positives), while Recall as TP/TP+FN (FN: False Negatives). Precision and Recall are alternatively called as Positive Predictive Value (PPV) and Sensitivity, respectively. We created a confusion matrix to better understand the image classification results (Fig. 5). The diagonal values in the matrix refer to the number of correctly identified pixels. The ratio of a total number of correctly identified pixels to the total number of considered pixels gives the classification's overall accuracy. The overall accuracy metrics, however, is influenced by unbalanced and prominent classes. Kappa index is therefore used to compare the classifiers. Kappa index of Agreement has been widely used in assessing the classifier's performance in remote sensing domain. Kappa value provides the information on the classifier as better or worse than random assignment of classes. The Kappa value of CNN classifier essentially suggests that the classifier is 81 percent better than random assignment of classes.

Table 3: Class-wise comparison of results from different classifiers.

| Classifiers | CNN | ANN | RF | SVM |
|---|---|---|---|---|
| Classes | F1-score | | | |
| LCZ 2 | 0.87 | 0.8 | 0.91 | 0 |
| LCZ 3 | 0.92 | 0.79 | 0.99 | 0.57 |
| LCZ 4 | 0.8 | 0.4 | 0.53 | 0.21 |
| LCZ 5 | 0.61 | 0.58 | 0.28 | 0 |
| LCZ 8 | 0.57 | 0.09 | 0.66 | 0 |
| LCZ 9 | 0.69 | 0.65 | 0.62 | 0.12 |
| LCZ 10 | 0.77 | 0.74 | 0.76 | 0.03 |
| LCZ A | 0.98 | 0.99 | 0.99 | 0.98 |
| LCZ B | 0.85 | 0.91 | 0.94 | 0.84 |
| LCZ C | 0.72 | 0.68 | 0.76 | 0.22 |
| LCZ D | 0.94 | 0.9 | 0.98 | 0.74 |
| LCZ E | 0.81 | 0.28 | 0.33 | 0 |
| LCZ F | 0.86 | 0.84 | 0.94 | 0.62 |
| LCZ G | 0.98 | 0.89 | 0.94 | 0.74 |
| | | | | |
| O.A. | 0.82 | 0.72 | 0.74 | 0.49 |
| Kappa | 0.81 | 0.69 | 0.72 | 0.44 |

As evident from Table 3, CNN classifier outperforms the pixel-based methods by a considerable margin. CNN is followed by RF, ANN, and SVM; with an overall accuracy of 74, 72, and 49 percent respectively. CNN has shown superior classification capability where per class F1-score is more than 0.70, except for the three classes. The score (0.57-0.92) of LCZ 3 stands out from the rest of the built classes in all classifiers, which shows the clear distinction of the pixel values of compact low-rise built form among all the built classes. Similarly, LCZ 2 shows consistency in performance in at least three of the classifiers. LCZ A and LCZ G, which represent dense trees and water respectively, are detected by all the classifiers with near cent percent accuracy. As discussed in the earlier text, CNN benefits from spatial and spectral variety, hence the built classes, which are studied in context with openness and vegetation (LCZ 4, 5, 9 and 10), have been better represented by the CNN classifier. The LCZ 8 class provides the least accurate predictions among the built classes, which is partially due to the presence of built structures comprising of large roof-areas in warehouses, low-rise structures, and homogeneity in pixel value distribution as in LCZ E and LCZ F classes. The RF classifier shows an affinity to LCZ E, where a fraction of almost every class is misidentified as impervious surfaces. LCZ 5 is interpreted as LCZ 4 in most cases due to the similarity in form and texture pattern and the probable inconsistency in the interpretation of both the classes while preparing the training samples. Among the natural classes, LCZ C is the most misclassified class. The main reason can be attributed to the use of Google high-resolution imagery used to visually select samples, while Sentinel imagery was used to create original training dataset. It may have led to the inclusion of patches which are prominent in High res imagery but not discernible to the classifier in mid Resolution Sentinel imagery.

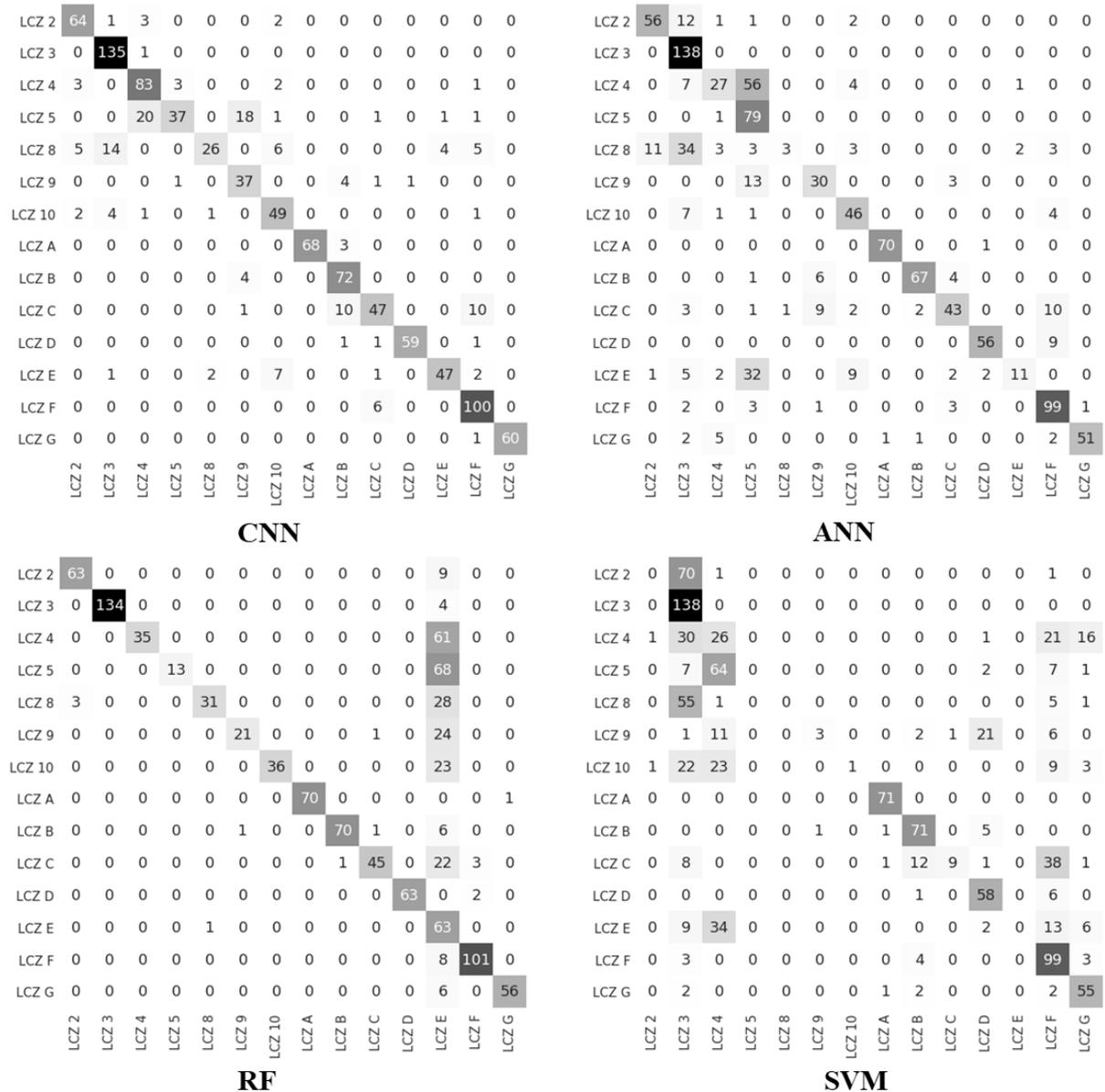

**Figure 5**: Confusion matrix showing the performance of classifiers.

The Sentinel 2B satellite data for Mumbai after clipping the imagery to city extents consists of 4192 x 2192 x 10 pixels. A sliding window of 11x11 pixels with a stride of 1 pixel is passed through the CNN trained classifier, the obtained result for every patch is represented by the center pixel of the patch. For pixel-based classifiers such as ANN, RF, and SVM; each pixel with 1x10 dimension is passed through the respective classifiers to obtain predictions. The spatial regulation on the predicted map is applied with the sliding window MV approach (Fig. 4). The final classification map is prepared with the use of different Python modules such as Rasterio [72] and Geospatial Data Abstraction Library (GDAL) [73]. The prediction map generated from CNN classifier is shown in Fig. 6.

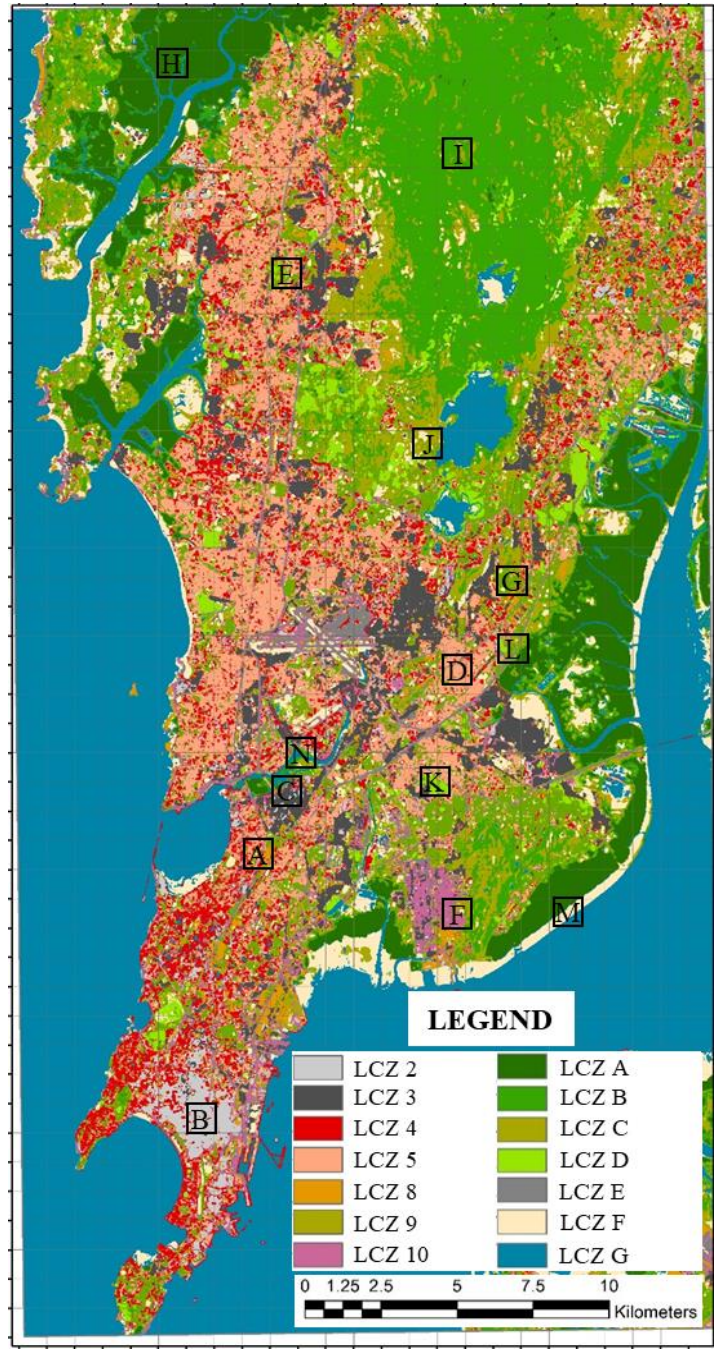

**Figure 6**: Classification map produced by CNN classifier (Additional maps in Fig. 7, 8, and 9).

The CNN based classifier generates the classification map at 10 m resolution. The performance of the classifier is excellent in demarcating different human-made and natural textures at fine spatial resolution. It is surprising that only a few points of training data can create a relatively high-resolution classification map (Fig. 6) which includes complex urban classes. Figure. 7, 8, and 9 show a zoomed version of the classification map showing each of the 14 classes. High-rise built structures (LCZ 4) do not form a continuous stretch, which is evident from the classification map. These structures are present in the form of towers in the midst of mid-rise colonies and structures. High rise buildings can be seen (in Red) as scattered in Fig. 7 A and 7 B in the midst of open and compact mid-rise built form (LCZ 2 and 5). The compact mid-rise built (LCZ 2) character extends to form a contiguous stretch (Fig. 7 B). The classification accuracy for Compact low-rise (LCZ 3), which is mostly represented by squatter settlements is highest among other built classes. Scattered slum settlements even having different forms and sizes are accurately demarcated in the city (Fig. 7 C). Figure 7 E shows the few occurrences of the LCZ 8 in the form of warehouses (in orange). Heavy industries (LCZ 10) can be visually identified by pipelines, tankers, and machinery in Google imagery. However, the accuracy of the classifier in detecting these features is comparatively lower. Figure 7 F shows LCZ 10 (in magenta) along with the misclassified categories. The sparse built cover (LCZ 9) includes an arrangement of built structures with large vegetation cover (Fig. 8 G). Natural classes such as LCZ A (Fig. 8 H) and LCZ B (Fig. 8 I), LCZ C (Fig. 8 J), and LCZ D (Fig. 8 K) are easily delineated by the CNN classifier. Coastal areas are highlighted (Fig. 9 M) which shows the presence of bare sand and soil (LCZ F). Figure 8 L shows road networks (LCZ E) (in grey) delineated by the classifier.

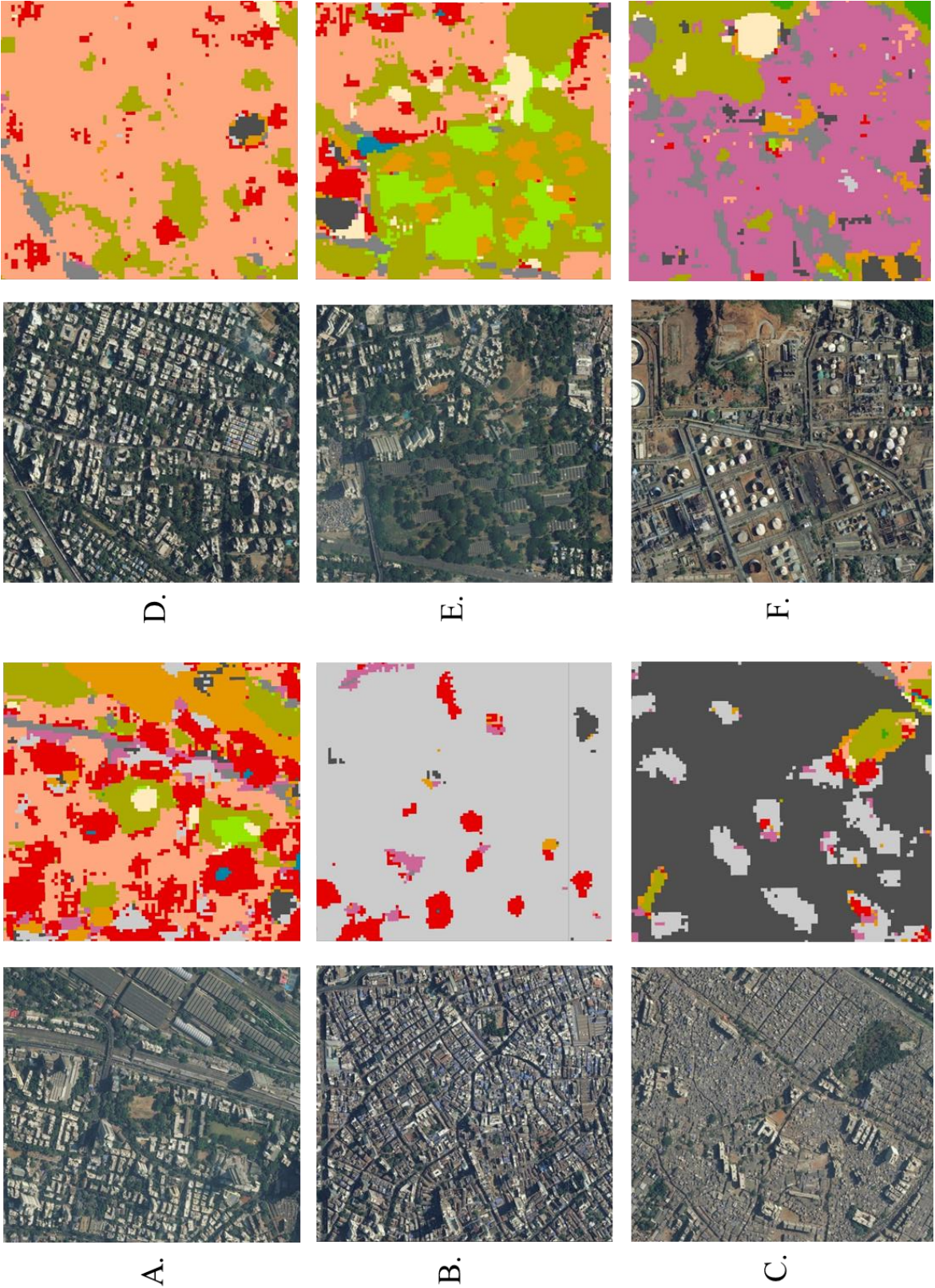

**Figure 7**: Zoomed in areas showing Google imagery and corresponding CNN classification performed on Sentinel 2B imagery (Part A).

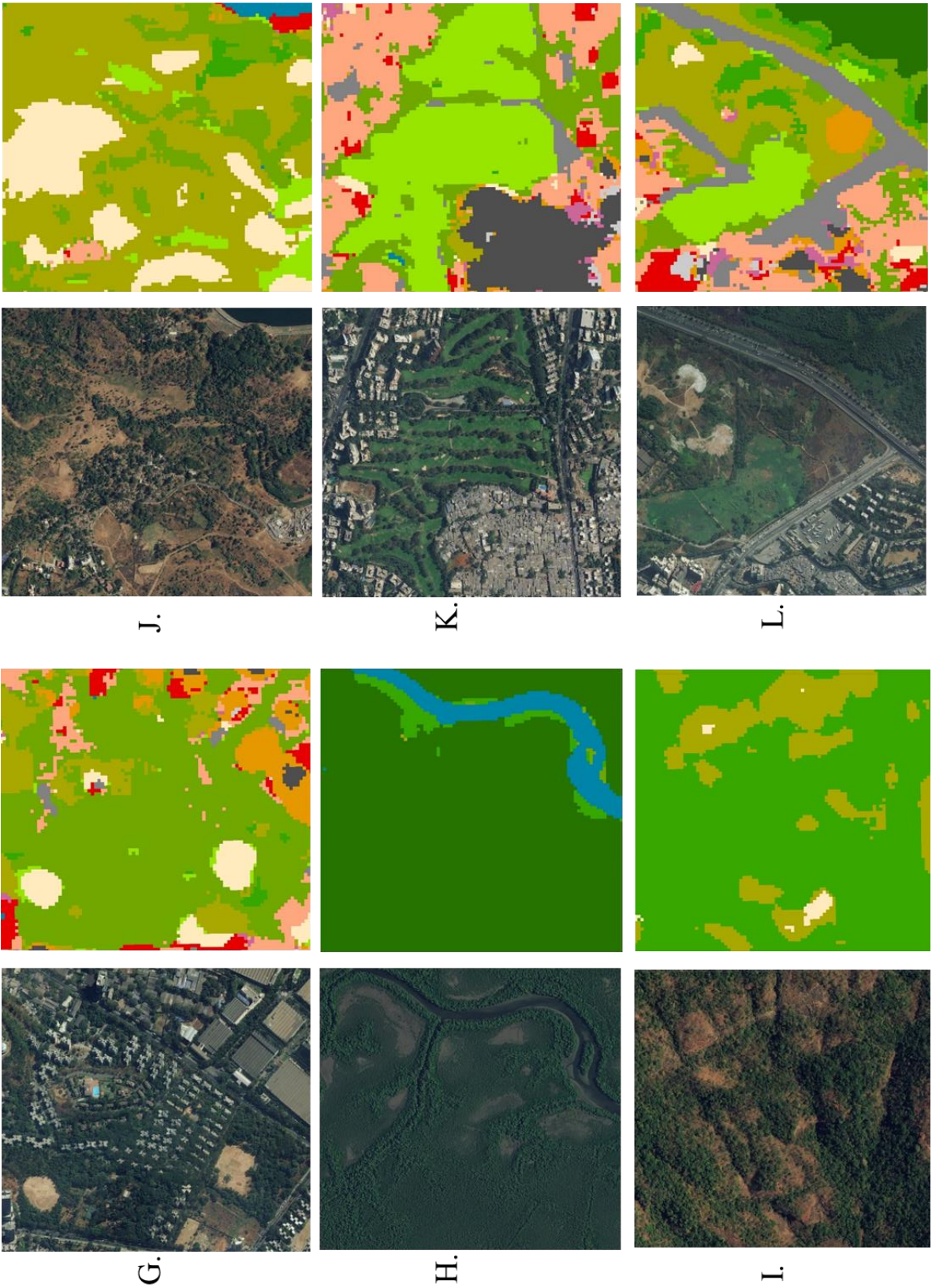

**Figure 8**: Zoomed in areas showing Google imagery and corresponding CNN classification performed on Sentinel 2B imagery (Part B).

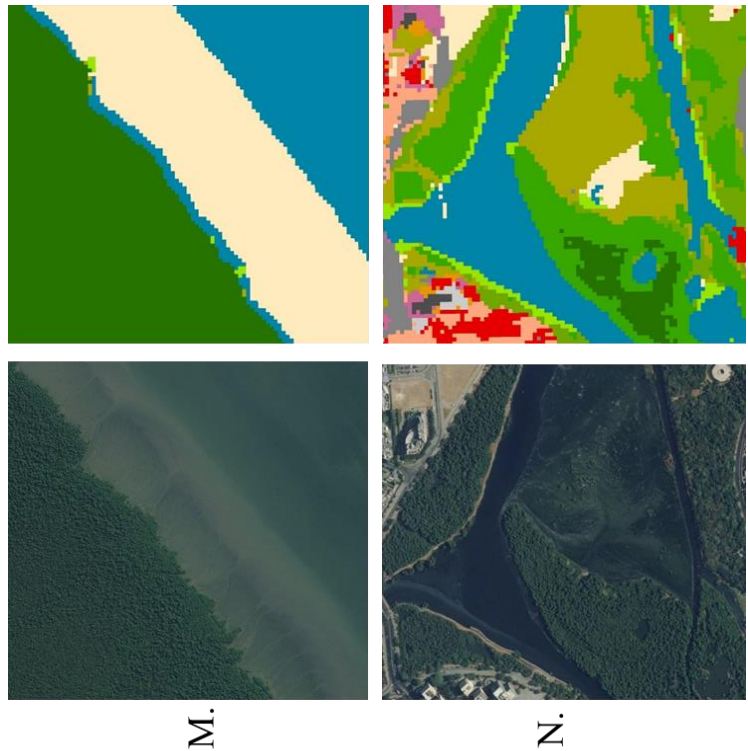

**Figure 9**: Zoomed in areas showing Google imagery and corresponding CNN classification performed on Sentinel 2B imagery (Part C).

The CNN-based classifier produced excellent results in classifying 14 classes. However, some noticeable errors in the classification map can be seen. Common urban features and structures such as flyovers, bridges, and steep ridges are misinterpreted as high-rise (LCZ 4) class. Further, some of the pixels belonging to the LCZ 4 are misclassified as water (LCZ G) due to the inability of the classifier to detect the structure beneath the shadow cast by the tall buildings. The pixels representing heavy industry (LCZ 10) class are present at various locations in the city, while it exists only in the southern part of the city. It is due to the fact that industry-like physical features are detected by the classifier in multiple locations such as metro/railway stations, airport buildings, Warehouses, flyovers, and subways.

## 6. Discussions

### 6.1. Conversion of classification map as per WUDAPT protocol

LCZ mapping process follows WUDAPT protocol, which allows local experts from all over the world to create consistent LCZ maps for different cities. For the universal applicability of the procedure, easier data availability, along with computationally and fiscally inexpensive software requirements have been given utmost focus. Several studies have modified the protocol to include open EO datasets such as SAR [37], [43], ASTER [42], and proprietary VHR datasets [74] to prepare LCZ maps to gain accuracy and better class delineation. Further the inclusion of OBIA techniques [74] and rule-based analysis from GIS datasets [38], [39] have also been used in preparation of such maps, which have shown comparatively better classification results [36]. However, these additional techniques and methods require advanced image analysis skills and software knowledge which adds to the complexity for an inexperienced user and hence creates a lack of reproducibility by experts in different cities [9].

The proposed methodology used in this study is different from WUDAPT protocol in that it (a) takes into account 10 band Sentinel Imagery as opposed to Landsat imagery, (b) considers point-based samples instead of polygons to overcome the difficulty in sample creation in cities with large horizontal heterogeneity, and (c) utilizes native spatio-contextual classifier such as CNN and modified machine learning classifiers to prepare classification maps. While openly available Sentinel dataset and selection of point-based samples can be easily implemented as a part of LCZ maps preparation, the implementation of Contextual classifiers such as CNN requires experience and training, which challenges the original purpose of WUDAPT protocol. Studies such as [45] detail the need to integrate spatial information as part of the WUDAPT classification process, especially in cities with spatial heterogeneity. However, such methods are relatively difficult to replicate without proper documentation regarding the implementation of such tools. More studies are required to explore the spatial-spectral domain in LCZ classification.

According to [9], the spatial resolution of LCZ maps at 100-150m is optimal to fulfill the original purpose of the LCZ maps preparation, i.e., to classify Urban Heat Island (UHI) observation sites and to represent the climate zones at the local scale. The discussed methodology generates the classification output as 10m resolution, which in LCZ classification method is considered too high due to the creation of a large number of isolated pixels. In this study, the problem of isolated pixels and patches in classification results at a higher resolution is to a great extent solved with the help of spatio-contextual classifiers such as CNN and modified machine learning image classification methods. Further, to decrease the granularity and to create homogeneous zones, post-classification filters [36], [75] can be applied in accordance to [9]. Following this approach, the classification map can be effectively converted to a LCZ map for the city while also being relevant to various other use cases.

## 6.2. Relevance of high-resolution LULC maps in urban studies.

Apart from urban climate-oriented studies, LULC mapping using LCZ scheme has relevance to different domains and subjects of societal benefit. The high-resolution classification maps can act as a remote sensing based proxy for micro-level zone delineation which may assist in urban disaster risk management (DRM) and vulnerability mapping [76]. The demographic characteristics and socioeconomic status of urban areas are closely related to urban morphology [77], [78]. The areas represented as LCZ 3 (Fig. 7 C) mostly represent slum settlements, the location and growth of which can be efficiently mapped and empirical connections can be drawn with indicators such as population density, requirements of various services, and potential health risks. As the classification maps are prepared on openly available Sentinel dataset, such assessments can be carried out at the regional or national level to frame policies [77], [79], [80]. Availability of urban open and green spaces have been extensively studied as a recreational potential and wellbeing with the help of various remote sensing techniques [81]–[83]. The classification results obtained from the proposed approach clearly demarcate the location of open playgrounds and green areas (Fig. 7 A, C and Fig. 8 G), even when these classes are enclosed by other built or natural classes. Micro-level assessments regarding the accessibility to open spaces and calculation of natural and built character can greatly benefit from such classification maps. In a similar way, image classification strategy discussed in the study may improve the classification accuracy to detect impervious surfaces [84] and different natural classes, which help in estimation of water runoff [85] and assessment of floods [86]. Overall, such classification maps will hold great relevance to urban management authorities who devise land use and site planning regulations [87] and monitor urban change detection and haphazard development.

## 6.3. Limitations of the proposed methodology

This study discussed the application of CNN in detail. However, the preparation of the CNN model to fit every use case is a laborious task. The hyperparameter tuning includes a series of trial and error ranging from the size of the input to the selection of activation functions. The size of the input patch in satellite imagery classification tasks is a tradeoff between the introduction of neighborhood spatial heterogeneity and model accuracy. Finding a sweet spot between the two factors is difficult. Further, CNN models require a high computational cost. Hence the proper evaluation of the requirements and purpose is necessary before deciding to use CNN based classification methods. Further studies are required to evaluate the model performance using selected imagery bands and fusion of other SAR and multispectral satellite datasets.

Integration of spatial information in machine learning classifiers can be executed with the help of different methods; this study utilized one of the many such approaches. Similar to the application of CNN, Majority Voting (MV) at neighborhood for each central pixel creates a huge computational load. Such issues create major bottlenecks during the implementation process. The evaluation of other spatial-spectral methods can be checked to select the optimal approach. Further, the MV approach considers the fixed neighborhood size of 11x11 pixels, which is kept equal to the spatial input to the CNN model. The optimum neighborhood size can be determined with further reiterations.

## 7. Conclusions

In this study, we utilized state-of-the-art deep learning techniques such as CNN and open satellite data sources to prepare a classification map as per the LCZ scheme. While high-resolution datasets have been used to study the effectiveness of DL algorithms, studies demonstrating the applicability of such techniques in mid-resolution satellite imagery are scarce. Surprisingly, the performance of the created model is exceptional, given the inherent complexity in urban texture mapping in a city like Mumbai and the relatively coarser resolution of the Sentinel dataset. The process of creating such maps includes point-based training data generation, which is achieved with the help of Google Earth Pro software. The creation of training samples can be crowdsourced to achieve generalizability and more local expertise in the selection of LCZ classes. This study provides a novel end to end approach to produce classification maps with better accuracy and a relatively more straightforward method. The approach differs from other imagery-based LCZ mapping methods as it considers the spatial and spectral variations by comparing the native spatial-spectral classifier with spectral-only ML classifiers modified to include spatial information.

Seasonal variations can be further studied to prepare a robust classification map. The presence of natural classes, especially low plants, which indicate open fields with grass cover and bush and scrubs vary with the seasons. Multitemporal classification maps may give insight into the change in natural characters and thermal profiles in the city throughout the year [35], [75]. The creation of an integrated GIS tool to accomplish the samples generation and CNN model building may help in rapid prototyping and quick results. Such methods can be transformed into larger studies including various cities and regions with relative ease. In recent years, much research and development have been focused on establishing newer DL algorithms to solve a variety of problems in different domains. Translating such efforts in remote sensing may help in uncovering new frontiers. For example, studies have been conducted which measure poverty and other social indicators through satellite imagery [88], [89]. Likewise, the delineation of urban features with openly available satellite imagery may assist in rapidly changing urbanscapes in developing and underdeveloped nations.

**Acknowledgments**

The authors would like to thank the Ministry of Human Resource Development (MHRD), India and Industrial Research and Consultancy Centre (IRCC), IIT Bombay for funding this study under the grant titled Frontier Areas of Science and Technology (FAST), Centre of Excellence in Urban Science and Engineering (grant number 14MHRD005).